# Phonon-interference resonance effects in nanoparticles embedded in a matrix

Lei Feng[1], Takuma Shiga[1], Haoxue Han[2], Shenghong Ju[1], Yuriy A. Kosevich[3†], and Junichiro Shiomi[1,4*]

[1]*Department of Mechanical Engineering, The University of Tokyo, 7-3-1 Hongo, Bunkyo, Tokyo 113-8656, Japan*

[2]*Theoretische Physikalishe Chemie, Eduard-Zintl-Institut für Anorganische und Physukalische Chemie, Technische Universität Darmstadt, Alarich-Weiss-Straße 4, 64287 Darmstadt, Germany*

[3]*Semenov Institute of Chemical Physics, Russian Academy of Sciences, Kosygin Str. 4, Moscow 119991, Russia*

[4]*Center for Materials research by Information Integration, National Institute for Materials Science, 1-2-1 Sengen, Tsukuba, Ibaraki 305-0047, Japan*

†Email: yukosevich@gmail.com

*E-mail: shiomi@photon.t.u-tokyo.ac.jp

We report an unambiguous phonon resonance effect originating from germanium nanoparticles embedded in silicon matrix. Our approach features the combination of phonon wave-packet method with atomistic dynamics and finite element method rooted in continuum theory. We find that multimodal phonon resonance, caused by destructive interference of coherent lattice waves propagating through and around the nanoparticle, gives rise to sharp and significant transmittance dips, blocking the lower-end frequency range of phonon transport that is hardly diminished by other nanostructures. The resonance is sensitive to the phonon coherent length, where the finiteness of the wave packet width weakens the transmittance dip even when coherent length is longer than the particle diameter. Further strengthening of transmittance dips are possible by arraying multiple nanoparticles that gives rise to the collective vibrational mode. Finally, it is demonstrated that these resonance effects can significantly reduce thermal conductance in the lower-end frequency range.

Controllability of thermal transport in materials is highly important in order to meet the technological needs to dissipate, store, or convert thermal energy. For instance, the suppression of thermal transport leading to low thermal conductivity is beneficial for thermoelectric materials [1]. The thermal transport in common crystalline materials is a highly multiscale phenomenon where thermal phonons with a broad range from sub- to tens of terahertz (THz) contribute [2,3]. Therefore hierarchically-structured materials such as those combining the grain boundaries and impurities capable of annihilating broad range of phonons are comparatively effective [4,5]. For further reduction of thermal conductivity, the key is to inhibit transport of phonons with the *lower-end frequencies* (from sub THz to a few THz) because they tunnel through the interface (grain boundary) since the transmittance asymptotically approaches unity as frequency decreases [6]. The exact critical frequency below which the tunability becomes impacting depends on the material, but for instance a recent study on crystal-amorphous silicon (Si) nanocomposite has shown that phonons with frequency below a few THz still propagate and contribute to a large fraction of the remaining thermal transport [7]. Such significance of phonons with the *lower-end frequencies* should be applicable in general for nanostructured crystalline materials with low thermal conductivity [8,9].

A widely explored approach to impede low frequency phonons is to construct a phononic crystal, which inhibits propagation of phonons within certain frequency range as a consequence of interference of phonon waves reflected at the periodic structures [10]. A challenge from practical viewpoint lies in the necessity to pattern the periodic structures at the nanoscale such as the epitaxial superlattices. Although top-down nanofabrication (such as holes) with length scale of ~100 nm is possible [11-14], the target phonon frequency would be limited to the order of gigahertz, which has negligible contribution to thermal transport at room temperature due to the small density of states.

One way to introduce phonon interference without having to construct spatially periodic structures is to exploit local resonance. This has been theoretically demonstrated in various systems with the "added-structures" such as nanowires and thin films with pillars erected on the surface [15-19], and a solid interface with embedded defect-atom arrays [20,21]. The effect of local resonance on reflection enhancement can be related with destructive interference of different phonon paths in real space (through and around the local resonator), and results in flattening of phonon bands or in total reflection of phonons at certain frequencies [6,15,20,21]. However, to impact phonons with the *lower-end frequencies*, the above "added-structures" need to be built at the nanoscale, and thus would still be extremely challenging.

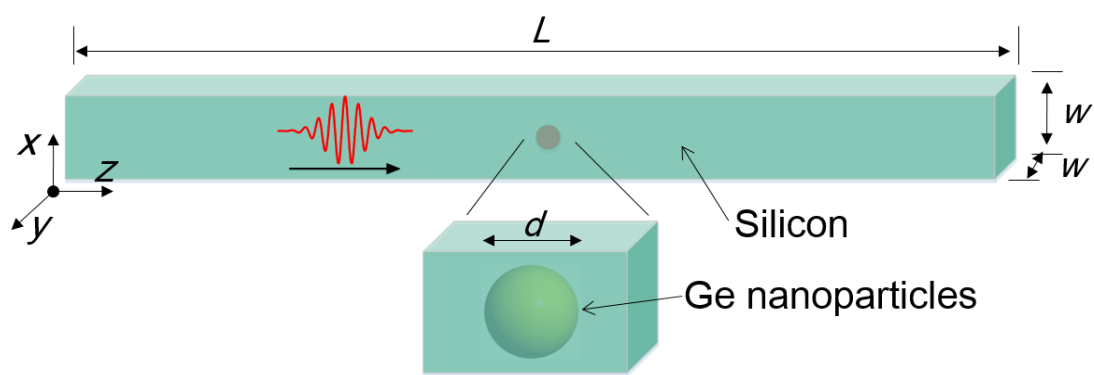

**Fig. 1** Configuration of phonon wave-packet (PWP) simulation. $L$ denotes the length of simulation domain, $w$ is the side length of the square cross section, and $d$ is the diameter of GeNP centered in the box.

In this Rapid Communication, we explore the possibility to introduce the local resonance in a practical system, where the coherently embedded germanium nanoparticles (GeNPs) in Si matrix are considered as nano-oscillators interacting with lattice waves [15] and similar structures have been fabricated in Refs. [22,23]. We conduct polarization-wise phonon wave-packet (PWP) simulations [24-26] based on molecular dynamics (MD) of both longitudinal and transverse acoustic (LA and TA) waves to retrieve the resonance frequencies, transmittance, and associated vibrational mode of the GeNP and highlight the impact of coherence length on resonance effect. A representative configuration of the PWP simulation is depicted in Fig. 1 and its details are in Supplementary Materials [27]. We ensure the same area fraction ($\pi d^2/4w^2$) of the spherical GeNPs when varying their diameters $d$ and side lengths of the square cross section $w$. The relation of the local resonance in GeNP with the classical problem of dynamic deformation of an elastic particle embedded in a matrix is highlighted through the analysis of vibrational eigenstates with finite element method (FEM) based on continuum theory. Possibilities to enhance resonance reflection is discussed by varying coherence length of PWP and forming an array of GeNPs for collective modes. Finally impact of the resonance effect on thermal transport is quantified by atomistic Green's function (AGF) method [28,29] calculating frequency $\omega$ dependent spectral thermal conductance $G(\omega)$.

The transmittance $\alpha(\omega)$ of LA and TA PWPs for a single spherical GeNP are shown in Figs. 2(a) and (b), respectively. It shows that $\alpha(\omega)$ has several local transmittance minima, while the base-line gradually decreases as frequency increases. Among the local minima, large transmittance dips are clearly observed in a few THz range for both LA and TA phonons. To identify their origins, we retrieve time-evolution of the center of mass (COM) of GeNP ($d$=1.1 nm) at the frequency of minimum transmittance. As the LA PWP passes through the GeNP, the vibrational amplitude of COM transiently increases and then decreases. The COM remains vibrating even after PWP has passed away, with temporal period corresponding to the resonant frequency $\omega_R$=1.89 THz, which indicates the resonance with the incident phonon. Following the polarization of the LA PWP, the GeNP vibrates only along the $z$-axis, i.e., the resonating GeNP eigenmode is a translational mode with "rattling" motion, as sketched in the inset of Fig. 2(a). This resonant mode was found to be the same for GeNPs with other diameters

[27].

For TA PWP, both the *x*- and *y*-coordinates of the COM exhibit sinusoidal vibrations with $\omega_R$=2.05 THz for *d*=1.1 nm. In this case, vibrations of GeNP take place in both *x*- and *y*-axes following the eigenvectors of the TA phonon. This results in rotational motion in the *x*-*y* plane as sketched in the inset of Fig. 2(b), which here is termed as "libration".

FEM analysis computing the vibrational eigenfrequencies of embedded GeNPs was conducted by COMSOL Multiphysics® v5.2a software. Here, Young's modulus (100 GPa) and Poisson ratio (0.335) of materials are calculated from lattice dynamics [30] using the same potential in PWP simulation for consistency. By adopting the same configuration as that of the PWP simulation, we identify the eigenfrequencies of the GeNP whose eigenmodes match with the motions observed in the PWP simulation. In Figs. 2(c) and (d), the diameter dependences of the eigenfrequencies for LA and TA modes are compared with that of resonant frequencies obtained from PWP simulation. The eigenfrequencies agree well with the resonant frequencies, although the small discrepancy slightly grows as *d* decreases since the shape of GeNP deviates from an ideal sphere. The frequency linearly scales with inverse diameter, i.e. $\omega_R d$ is invariant for the same mode under the same area fraction, which is a reminiscent of the frequency-spectra scaling law of the quasimacroscopic-acoustics origin, see also [31]. In this linear dispersion regime, this can be also written in terms of the central wavelength of PWP *λ* as *λ*≈4*d* and *λ*≈2.6*d* for LA and TA PWPs, respectively.

Note that the transverse periodicity of GeNPs imposed naturally in our PWP simulation (with one GeNP per transverse supercell) is not necessary for the current resonance effect to take place as the resonant frequency and transmittance dip are found to be similar even by randomly displacing the GeNPs, i.e. breaking the periodicity [27]. This confirms the advantages of such local resonance over those requires rigorous global periodicity. Also, the transverse periodicity leads to different number densities of GeNPs for LA and TA modes manifesting in slightly different resonant frequencies, which are otherwise the same for an isolated GeNP.

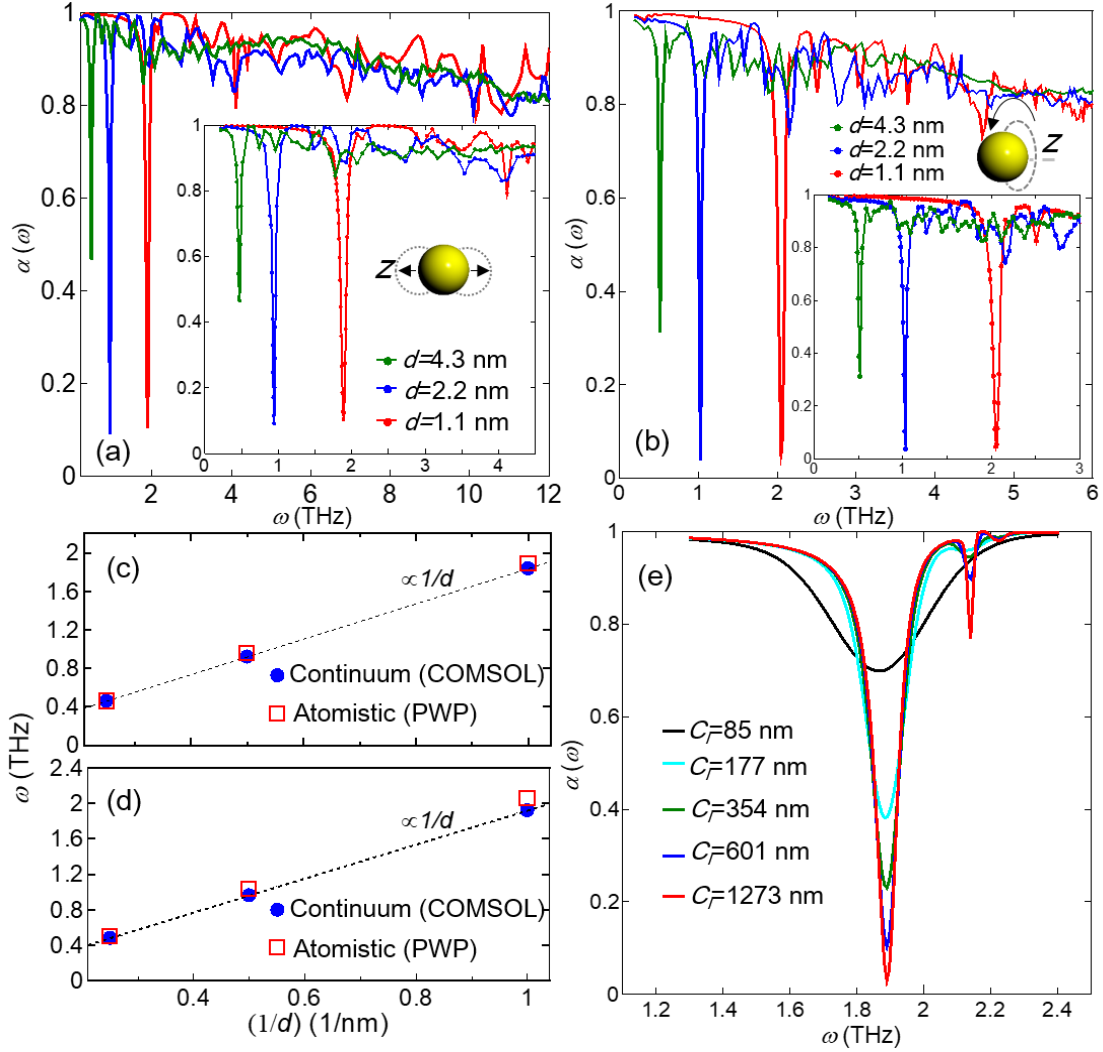

**Fig. 2** (a) and (b) Frequency-dependent transmittance $\alpha(\omega)$ calculated by PWP simulations for longitudinal acoustic (LA) and transverse acoustic (TA) phonons with $d$=1.1, 2.2, and 4.3 nm. Inset schematics show the motions of GeNP (*rattling* or *libration*). (c) and (d) Diameter-dependent resonant frequencies for LA and TA phonons (open red squares). Blue filled circles are eigenfrequencies calculated from continuum theory. The dotted lines denote the inverse $d$-dependence, $1/d$. (e) Variation of the LA-transmittance dip with different coherence lengths $C_l$=85, 177, 354, 601 and 1273 nm with $d$=1.1 nm.

We highlight the effect of the coherent length $C_l$ on resonance as $C_l$ can be easily tuned in our PWP simulation. In reality, it takes a finite value determined by phonon scattering due to anharmoniciy, impurity, and/or defects, and thus, depends on the actual system and temperature. Figure 2(e) summarizes the change in transmittance dip for LA PWP ($d$=1.1 nm) by varying $C_l$ as 85, 177, 354, 601 and 1273 nm. It is seen that, by increasing $C_l$, the depth and width of the dip increases and decreases, respectively, and eventually would lead to a complete reflection at the resonance frequency for infinite $C_l$ originated from the destructive interference. In case of finite $C_l$, as size of PWP becomes shorter and range of frequency components becomes broader, the transmittance dip, that is given by the convolution of PWP and the resonant mode, is no longer zero at the resonance frequency [21,27]. An important observation here is that the weakening of resonance manifests for coherent length that is much larger than the particle size. For instance, the magnitude of the transmission dip was reduced by 40% even though $C_l$ is more than 100 times larger than $d$. There have been many works reported recently aiming to establish phononic materials with global or local phonon interference, and the usual challenge has been to reduce the structure sizes below the coherence length. However, the present finding indicates that the structure needs to be orders-of-magnitude smaller than the phonon coherent length for the interference to give the impact anticipated from the plane-wave-based analysis. Therefore, we expect that the resonance effect would be largely constrained in reality unless very small structures such as the current nanoparticles are used.

In addition, the large $C_l$ calculation finds the presence of a secondary dip (Fig. 2 (e)), at a frequency higher than the fundamental one, which originates from resonant squeeze mode of GeNP [27]. For the rest of the transmittance calculations, we adopt a fixed value of $C_l$=354 nm for all the frequencies except for those around the largest dips, with which dip width starts to saturate, and the computation is affordable. It should be noted here that $C_l$=354 nm is on the order of the phonon MFP of pure crystal Si at room temperature. As for the frequencies around the largest dip, $C_l$ was set to 550$d$ in case of $d$=1.1, 2.2 nm to the assure saturation, while in case of $d$=4.3 nm, $C_l$ was limited to 140$d$ due to limitation in computational resources.

Besides the largest transmittance dips, the presence of other smaller dips is also important for thermal transport. For instance a resonant dip at $\omega_{2R}$=4.12 THz is observed with $d$=1.1 nm in the inset of Fig. 2 (a), which is approximately two times larger than $\omega_R$=1.89 THz. The GeNP at $\omega_{2R}$ is found to resemble "rattling" motion at $\omega_R$ but with nearly one-order smaller amplitude, therefore we conclude that it is the second harmonics. The same relation is observed for other cases ($d$=2.2 nm: $\omega_R$=0.95 THz, $\omega_{2R}$=1.90 THz; $d$=4.3 nm: $\omega_R$=0.45 THz, $\omega_{2R}$=1.05 THz). At even higher frequencies, $\lambda$ becomes comparable or shorter than $d$, which is no longer in continuum regime but at atomistic scale, and the transmittance dips turn into fluctuations. From these, we identify three frequency regimes: (i) lowest frequency regime of the strongest resonance (the largest transmittance dip) with the fundamental modes, (ii) intermediate frequency regime of resonance with high-order harmonics, and (iii) highest frequency regime of atomistic-scale scattering.

The transmittance dip can be further enhanced by manipulating the inter-particle distance among multiple GeNPs to excite collective motions of them. For the demonstration, four spherical GeNPs ($d$=1.1 nm) are aligned along the $z$-axis with equal inter-particle distance $D$ to form an array with $D$=$d$, $2d$, $4d$, $5d$ and $8d$, of which two adjacent GeNPs are sketched in the inset of Fig. 3(a). Fig. 3(a) shows that except for $D$=$d$, depths of LA-transmittance dips are enhanced due to magnification of resonance by multiple GeNPs (similar for TA modes in [27]). For $D$=$2d$, the width becomes much larger than the single GeNP case. It is found that at the resonant frequency (the same frequency as single GeNP), four GeNPs exhibit out-of-phase vibration (adjacent GeNPs rattling oppositely along the $z$-axis) as sketched in Fig. 3(d)-(1). Recalling that $\lambda \approx 4d$ holds for the rattling mode, the out-of-phase collective vibration is understandable since each GeNP is located on the node of the phonon wave. Its robustness is further evidenced by the similarities among transmittance dips for $D$=$2d$, $4d$ and $8d$, which are integral multiples of $2d$. Furthermore, we have performed the FEM analysis for four GeNPs array with $D$=$2d$ and extracted four relevant eigenstates whose frequencies are close to the resonant frequencies as indicated in Fig. 3(a). The obtained vibrational modes are sketched in Fig. 3(d) in the order of ascending frequencies. Among the four modes, the out-of-phase vibration in Fig. 3(d)-(1) was observed in the PWP simulation because of the high receptivity, i.e. the agreement of eigenmodes between the PWP and collective resonance.

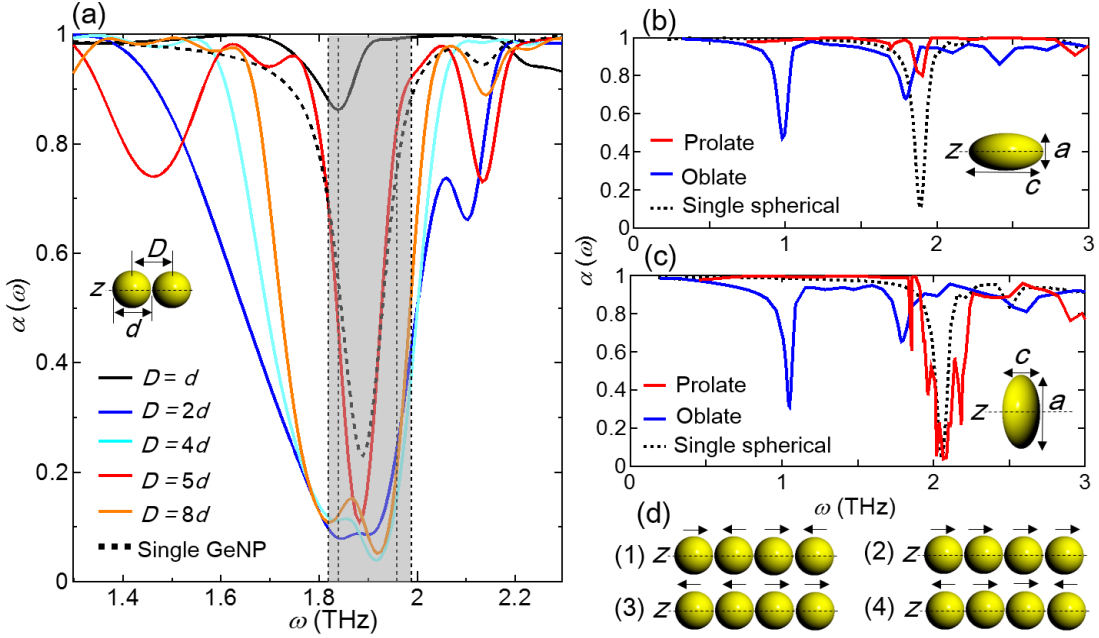

**Fig. 3** (a) Transmittance $\alpha(\omega)$ calculated from LA PWP simulation with four GeNPs array ($d$=1.1 nm) with different equal inter-particle distances $D$=$d$, $2d$, $4d$, $5d$ and $8d$. Inset: schematic for two GeNPs array along $z$-axis with $D$=$d$. Four vertical dot lines in shaded region denote four relevant eigenfrequencies for four GeNPs array ($D$=$2d$) calculated from continuum theory. (b) $\alpha(\omega)$ from LA PWP simulation with oblate and prolate types of ellipsoidal GeNPs. Inset: schematics for the prolate GeNP ($a$=$b$=1.1 nm, $c$=3.3 nm). (c) The same as (b), but for TA PWP. Inset: schematics for the oblate GeNP ($a$=$b$=3.3 nm, $c$=1.1 nm). (d) Sketches of eigenmotions corresponding to four eigenfrequencies in (a) in the order of ascending frequencies. The arrows indicate vibrational directions of each GeNP.

In the case of $D$=$5d$, the dip width is narrower due to the absence of collective resonance, although the depth is larger due to the enhanced reflection by multiple

GeNPs compared with the single GeNP case. It is interesting, however, that the additional dips on the sides (e.g. the dip in between 1.4 and 1.6 THz), whose origin is possibly related with the Fabry–Pérot-like interference in the finite-size Si matrix with multiple GeNPs, are the largest for this case. With $D=d$, the resonant frequency shifts and the dip depth are considerably reduced. In this case, the GeNPs are almost in contact and they can be considered as a single body consisting of four GeNPs. As an extreme case, we consider prolate-ellipsoidal GeNP ($c/a$=3:1, $a=b$=1.1 nm, $w_x=w_y$=2.2 nm) as shown in the inset of Fig. 3(b), and observe that the transmittance dip of LA PWP is significantly shallower than that of single spherical GeNP [Fig. 3(b)]. On the other hand, the transmittance dip of TA PWP becomes deeper and wider and displays noticeable spikes [Fig. 3(c)]. We also show the transmittance profiles for ellipsoidal GeNPs with oblate form ($a/c$=3:1, $a=b$=3.3 nm, $w_x=w_y$=4.3 nm) with the plane of longer side perpendicular to $z$-axis as shown in the inset of Fig. 3(c). The newly emerged dips at much lower frequencies around 1 THz for both LA and TA PWPs and changes of the overall profiles can be attributed to drastic variations in the effective area fraction or inter-particle distance.

Figures 4(a) and (b) show $G(\omega)$ at $T$=300 K with $d$=1.1 nm for $\Gamma$-point mode (subset modes with zero wavenumber in the $x$ and $y$ directions) and for all the modes (full Brillouin zone (BZ)), respectively. In the full BZ calculation, 10×10 uniform $k$-mesh was adopted to ensure convergence of $G(\omega)$. Significant reduction of $G(\omega)$ by single GeNP is observed in the *lower-end frequency* regime. The resonance dips can be seen more clearly in the $\Gamma$-point calculation because of smaller number of modes being superimposed. For instance conductance dips of single GeNP corresponding to the primary resonant frequency of LA and TA PWPs can be recognized, together with other harmonic-resonance dips. In the case of four GeNPs array ($D=2d$), the conductance dips are much deeper and wider as expected from the analysis above. The resonance effect is the most impacting at $\Gamma$-point mode in the frequency range of 1.5-2.2 THz with $d$=1.1 nm. Single GeNP gives 17.6% reduction of $G(\omega)$ purely due to resonance effects and the number increases to 41.5% in case of the array.

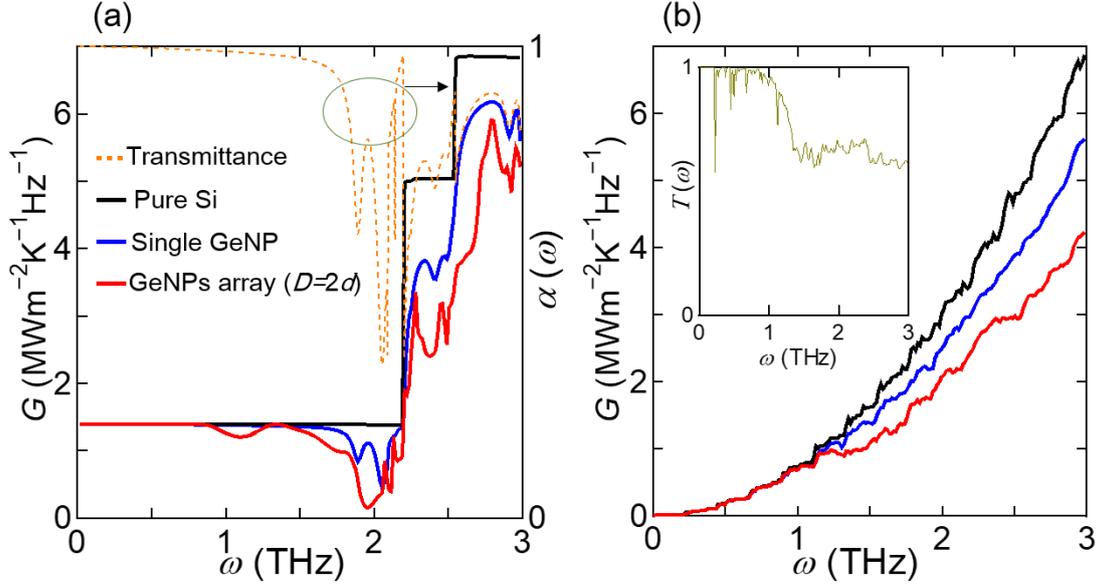

**Fig. 4** Spectral thermal conductance $G(\omega)$ at $T$ = 300 K by AGF (a) at Γ-point and (b) in full Brillouin zone. Pure Si without GeNP (black), single GeNP with $d$=1.1 nm (blue), and four GeNPs array with $D$=2$d$ (red). $\alpha(\omega)$ by AGF at Γ-point (orange dotted line) for single GeNP with $d$=1.1 nm is also superimposed in (a). Inset in (b): Transmittance spectrum $T(\omega)$ of four GeNPs array with respect to pure Si.

The resonant features become obscure in the full BZ calculation with dips of many modes with different wavevectors being superimposed, however, some of the features persist: the critical frequency above which the reduction becomes significant is about 1 THz, and four GeNPs array is evidently more effective than the single GeNP, whose effect is characterized by significant reduction in the transmittance spectrum $T(\omega)$ with respect to pure Si [inset in Fig. 4(b)]. Reduction of $G(\omega)$ for the full BZ calculation accounting for resonant contributions from other modes and non-resonance effects now becomes 15.8% for single GeNP and 33.7% for the array.

In summary, we report an unambiguous phonon-interference resonance effect originating from Ge nanoparticles embedded in Si crystal matrix. A spherical GeNP with a few nanometers in diameter resonates with acoustic phonon with *lower-end* *frequencies*. Finiteness of the coherence length leads to the broadening and shallowing of the transmittance dips, i.e. to the deterioration of the phonon-interference resonance effect unless the coherence length is two-orders-of-magnitude larger than the particle size. It thus highlights the necessity for structures at *true-nano-scale* as the present nano-particles when aiming to maximize the wave-interference effect in phononic structures in practice. The impact of resonance can be magnified by installing multiple layers of GeNPs due to the superposition of the resonant reflection and collective motion. Atomistic Green's function calculations accounting for all phonon modes in the Brillouin zone indicate that the resonance effects significantly reduce the thermal conductance in the *lower-end frequencies*. Narrow and tunable transmittance dips produced by embedded nanoparticles can be used for ultrasensitive measurements with phonon transmission spectra similar to ultrasensitive optical measurements in photonic crystals with embedded femtogram scale nanomechanical resonators [32].

This work was supported in part by CREST-JST and KAKENHI (Grand Nos. 16H04274, 15K17982). L.F. received support from Doctoral Student Special Incentives Program at Graduate School of Engineering, the University of Tokyo (SEUT RA) and Japan Society for the Promotion of Science (JSPS) Fellowship. Yu.A.K.'s visit to the University of Tokyo was supported by JSPS through Invitation Fellowships for Research in Japan. The computational resources for this work was provided by the Institute for Solid State Physics, the University of Tokyo.

[1] H. J. Goldsmid, *Introduction to Thermoelectricity*, pp. 9 (Phys Rev LettNano LettSpringer Berlin Heidelberg, Berlin, Heidelberg, 2016).

[2] D. A. Broido, M. Malorny, G. Birner, N. Mingo, and D. A. Stewart, Intrinsic lattice thermal conductivity of semiconductors from first principles, Applied Physics Letters **91**, 231922 (2007).

[3] D. Aketo, T. Shiga, and J. Shiomi, Scaling laws of cumulative thermal conductivity for short and long phonon mean free paths, Applied Physics Letters **105**, 131901 (2014).

[4] K. Biswas, J. He, I. D. Blum, C. I. Wu, T. P. Hogan, D. N. Seidman, V. P. Dravid, and M. G. Kanatzidis, High-performance bulk thermoelectrics with all-scale hierarchical architectures, Nature **489**, 414 (2012).

[5] T. Hori, G. Chen, and J. Shiomi, Thermal conductivity of bulk nanostructured lead telluride, Applied Physics Letters **104**, 021915 (2014).

[6] Y. A. Kosevich, Capillary phenomena and macroscopic dynamics of complex two-dimensional defects in crystals, Progress in Surface Science **55**, 1 (1997).

[7] Y. Zhou and M. Hu, Record Low Thermal Conductivity of Polycrystalline Si Nanowire: Breaking the Casimir Limit by Severe Suppression of Propagons, Nano Letters **16**, 6178 (2016).

[8] Y. Nakamura *et al.*, Anomalous reduction of thermal conductivity in coherent nanocrystal architecture for silicon thermoelectric material, Nano Energy **12**, 845 (2015).

[9] A. Miura, S. Zhou, T. Nozaki, and J. Shiomi, Crystalline–Amorphous Silicon Nanocomposites with Reduced Thermal Conductivity for Bulk Thermoelectrics, ACS Applied Materials & Interfaces **7**, 13484 (2015).

[10] S. Volz *et al.*, Nanophononics: state of the art and perspectives, The European Physical Journal B **89** (2016).

[11] J.-K. Yu, S. Mitrovic, D. Tham, J. Varghese, and J. R. Heath, Reduction of thermal conductivity in phononic nanomesh structures, Nature Nanotechnology **5**, 718 (2010).

[12] L. Yang, N. Yang, and B. Li, Extreme low thermal conductivity in nanoscale 3D Si phononic crystal with spherical pores, Nano Letters **14**, 1734 (2014).

[13] S. Alaie, D. F. Goettler, M. Su, Z. C. Leseman, C. M. Reinke, and I. El-Kady, Thermal transport in phononic crystals and the observation of coherent phonon scattering at room temperature, Nature Communications **6**, 7228 (2015).

[14] M. Nomura, Y. Kage, J. Nakagawa, T. Hori, J. Maire, J. Shiomi, R. Anufriev, D. Moser, and O. Paul, Impeded thermal transport in Si multiscale hierarchical architectures with phononic crystal nanostructures, Physical Review B **91**, 205422 (2015).

[15] Y. A. Kosevich, Multichannel propagation and scattering of phonons and photons in low-dimension nanostructures, Physics-Uspekhi **51**, 848 (2008).

[16] B. L. Davis and M. I. Hussein, Nanophononic metamaterial: thermal conductivity reduction by local resonance, Physical Review Letters **112**, 055505 (2014).

[17] D. Ma, H. Ding, H. Meng, L. Feng, Y. Wu, J. Shiomi, and N. Yang, Nano-cross-junction effect on phonon transport in silicon nanowire cages, Physical Review B **94**, 165434 (2016).

[18] S. Xiong, K. Saaskilahti, Y. A. Kosevich, H. Han, D. Donadio, and S. Volz, Blocking Phonon Transport by Structural Resonances in Alloy-Based Nanophononic Metamaterials Leads to Ultralow Thermal Conductivity, Physical Review Letters **117**, 025503 (2016).

[19] H. Honarvar, L. Yang, and M. I. Hussein, Thermal transport size effects in silicon membranes featuring nanopillars as local resonators, Applied Physics Letters **108**, 263101

(2016).

[20] H. Han, L. G. Potyomina, A. A. Darinskii, S. Volz, and Y. A. Kosevich, Phonon interference and thermal conductance reduction in atomic-scale metamaterials, Physical Review B **89**, 180301 (2014).

[21] Y. A. Kosevich, H. Han, L. G. Potyomina, A. N. Darinskii, and S. Volz, *Quodons in Mica*, pp. 247 (Springer International Publishing Switzerland, Switzerland, 2015).

[22] S. Yamasaka, Y. Nakamura, T. Ueda, S. Takeuchi, and A. Sakai, Phonon transport control by nanoarchitecture including epitaxial Ge nanodots for Si-based thermoelectric materials, Scientific Reports **5**, 14490 (2015).

[23] S. Yamasaka, K. Watanabe, S. Sakane, S. Takeuchi, A. Sakai, K. Sawano, and Y. Nakamura, Independent control of electrical and heat conduction by nanostructure designing for Si-based thermoelectric materials, Scientific Reports **6**, 22838 (2016).

[24] P. K. Schelling, S. R. Phillpot, and P. Keblinski, Phonon wave-packet dynamics at semiconductor interfaces by molecular-dynamics simulation, Applied Physics Letters **80**, 2484 (2002).

[25] K. Naoaki, Y. Takahiro, and W. Kazuyuki, Phonon wavepacket scattering dynamics in defective carbon nanotubes, Japanese Journal of Applied Physics **45**, L963 (2006).

[26] S.-H. Ju and X.-G. Liang, Investigation on interfacial thermal resistance and phonon scattering at twist boundary of silicon, Journal of Applied Physics **113**, 053513 (2013).

[27] Supplementary Materials, http://stacks.iop.org/1882-0786/8/i=7/a=071501.

[28] W. Zhang, T. S. Fisher, and N. Mingo, The Atomistic Green's Function Method: An Efficient Simulation Approach for Nanoscale Phonon Transport, Numerical Heat Transfer, Part B: Fundamentals **51**, 333 (2007).

[29] W. Zhang, T. S. Fisher, and N. Mingo, Simulation of Interfacial Phonon Transport in Si–Ge Heterostructures Using an Atomistic Green's Function Method, Journal of Heat Transfer **129**, 483 (2007).

[30] J. D. Gale and A. L. Rohl, The General Utility Lattice Program (GULP), Molecular Simulation **29**, 291 (2003).

[31] N. Combe and L. Saviot, Acoustic modes in metallic nanoparticles: Atomistic versus elasticity modeling, Physical Review B **80**, 035411 (2009).

[32] H. Zhang, C. Zeng, D. Chen, M. Li, Y. Wang, Q. Huang, X. Xiao, and J. Xia, Femtogram scale nanomechanical resonators embedded in a double-slot photonic crystal nanobeam cavity, Applied Physics Letters **108**, 051106 (2016).

# Supplementary Materials for

**Phonon-interference resonance effects in nanoparticles embedded in a matrix**

Lei Feng[1], Takuma Shiga[1], Haoxue Han[2], Shenghong Ju[1], Yuriy A. Kosevich[3†], and Junichiro Shiomi[1,4*]

[1]*Department of Mechanical Engineering, The University of Tokyo, 7-3-1 Hongo, Bunkyo, Tokyo 113-8656, Japan*

[2]*Theoretische Physikalishe Chemie, Eduard-Zintl-Institut für Anorganische und Physukalische Chemie, Technische Universität Darmstadt, Alarich-Weiss-Straße 4, 64287 Darmstadt, Germany*

[3]*Semenov Institute of Chemical Physics, Russian Academy of Sciences, Kosygin Str. 4, Moscow 119991, Russia*

[4]*Center for Materials research by Information Integration, National Institute for Materials Science, 1-2-1 Sengen, Tsukuba, Ibaraki 305-0047, Japan*

†Email: yukosevich@gmail.com

*E-mail: shiomi@photon.t.u-tokyo.ac.jp

1) Details of phonon wave-packet simulation (PWP)

A representative configuration of the PWP simulation is depicted in Fig. 1. The side lengths of the square cross section are set to be $w$=2.2, 4.4 and 8.6 nm, and the diameters $d$ of the spherical germanium nanoparticles (GeNPs) centered in the region are varied as $d$=1.1, 2.2 and 4.3 nm, respectively, which gives the same area fraction ($\pi d^2/4w^2$). A simulation domain with total length ($L$) of 1086 nm is used to allow transmittance calculation of long-wavelength phonons. The atomic displacement of PWP with branch $j$ and wave-vector center $\mathbf{k}_0$ is constructed by the following Eq. (S1) [S1-S3],

$$\mathbf{u}_j(kl,t) = \frac{A_j}{\sqrt{M_k}} \int d^3\mathbf{k}\, \mathbf{e}_j(k|\mathbf{k}) \exp[i\mathbf{k}\cdot(\mathbf{R}(l)-\mathbf{R_0})]\exp[-C_l^2\cdot(\mathbf{k}-\mathbf{k}_0)^2], \tag{S1}$$

where $\mathbf{u}_j(kl,t)$ and $\mathbf{R}(l)$ denote the displacement vector of the $k$-th atom in the $l$-th unit cell and the coordinates of the $l$-th unit cell, and $\mathbf{R}_0$ represents the center of PWP in real space. $\mathbf{e}_j(k|\mathbf{k})$ is the corresponding eigenvector for atom $k$ at given $\mathbf{k}$ and $j$. The amplitude $A_j$ was chosen as $2\times10^{-4}$ nm in order not to excite lattice anharmonicity. Here, coherence length (size of PWP) $C_l$ is set as 354 nm unless otherwise noted, much longer than the wavelength $\lambda_c$ of the central phonon of PWP ($C_l \gg \lambda_c$). A PWP is launched at a location far from the GeNP embedded region, and the frequency-dependent transmittance $\alpha(\omega)$ is calculated as the ratio of kinetic energies of the transmitted and incident waves. Note that there is lattice mismatch at Si/Ge boundary and possible distortion of Si matrix in reality, but we expect its effect is negligible based on previous study on phonon transmission and thermal conductance of strained Si/Ge interface [S4]. So we apply the same interatomic Stillinger-Weber (SW) potential of Si parameterizations [S5] to both Si and Ge atoms for convenience and only modify the mass for Ge ($m_{Ge}$=2.585$m_{Si}$). Moreover, the neglect of interatomic force field difference between Si and Ge atoms is justifiable because first-principles calculations have shown that the interatomic force constants (both harmonic and anharmonic) of Si and Ge crystals are mutually transferable [S6]. The actual simulation is performed using LAMMPS package [S7].

2) Vibrational modes analysis

We present the vibrational modes analysis of the GeNP ($d$=2.2 nm) with amplitude of phonon wave-packet being $2\times 10^{-5}$ nm from trajectories of Ge atoms corresponding to several frequencies of transmittance minima in Fig. S1.

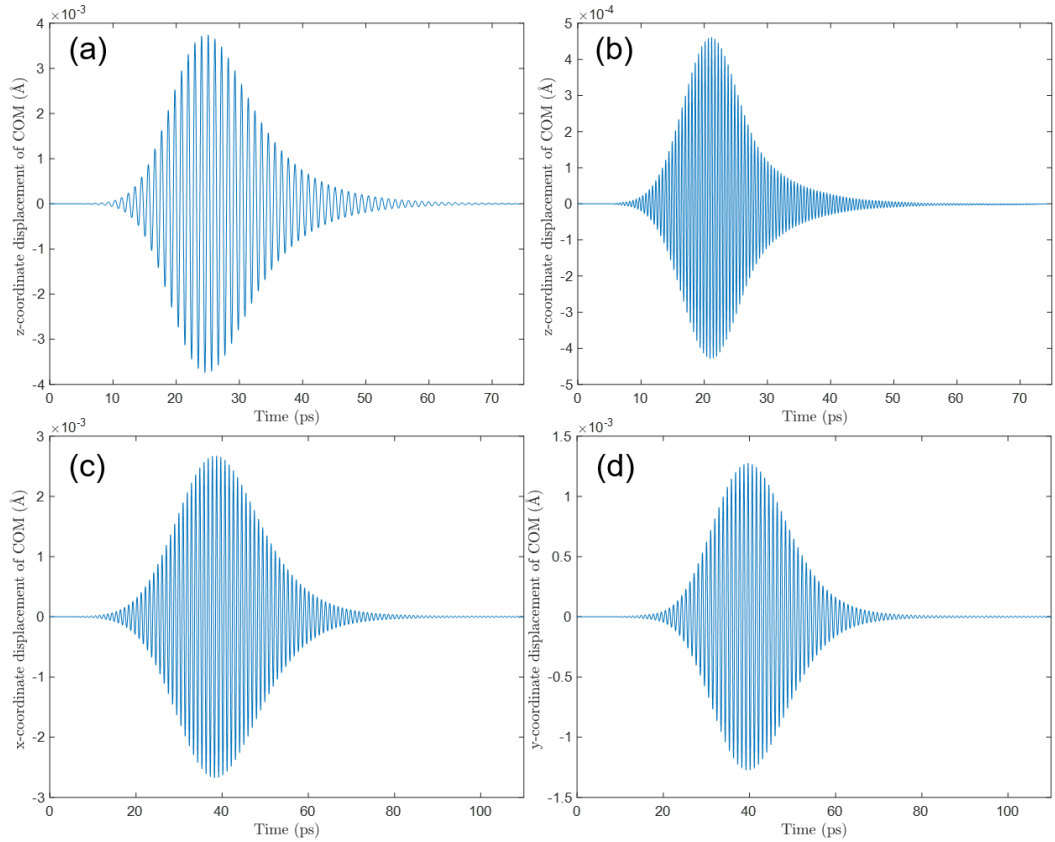

**Fig. S1** Time-evolution of $x$-, $y$-, or $z$-coordinate displacements of center of mass (COM) of single GeNP ($d$=2.2 nm) corresponding to several frequencies of transmittance minima. (a) The largest resonant transmittance dip in Fig. 2(a) for LA phonon. (b) Second

harmonics of the largest resonant transmittance dip in Fig. 2(a) for LA phonon. (c) The largest resonant transmittance dip in Fig. 2(b) for TA phonon (*x*-coordinate displacement). (d) The same as (c), but for *y*-coordinate displacement.

3) Weak dependence of resonant frequency and transmittance dip on the rigorous transverse periodicity

We show the weak dependence of resonant frequency and corresponding transmittance dip on the rigorous transverse periodicity by taking LA phonon with *d*=1.1 nm as example in Fig. S2. In PWP simulation, we change the original square cross section in Fig. 1 to rectangular cross section with $w_x$:$w_y$=2:1 containing two GeNPs in the inset (ii) of Fig. S2 to accommodate aperiodic variation. The GeNP on the right in the inset (ii) is then displaced by about 0.76 nm from its original periodic position as shown and denoted as "Aperiodic", while the inset (i) is "Periodic" and equivalent to original configuration in Fig. 1. The resulting largest transmittance dip is found to resemble that of the original configuration with a bit shallower depth, and resonant frequency shows slight blue-shift.

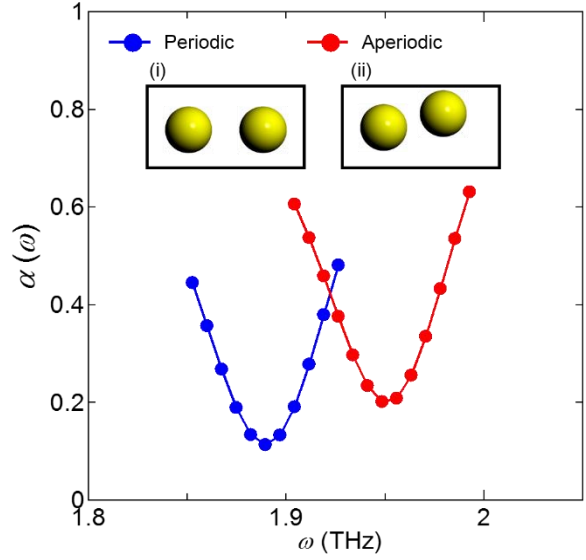

**Fig. S2** Comparison of $\alpha(\omega)$ between "Periodic" and "Aperiodic" configuration for LA PWP with $d$=1.1 nm calculated from PWP simulation. Inset (i): "Periodic" configuration. Inset (ii): "Aperiodic" configuration.

4) Resonant frequency dependence on area fraction and inter-particle distance

The resonant frequency of an embedded spherical object with transverse periodicity by a plane longitudinal wave is determined by a complex interplay among materials properties, area fraction of the object and inter-particle distance [S8]. We check several aspects which are relevant here.

Firstly, the resonant frequency is found to be dependent on the inter-particle distance. We demonstrate this by fixing $d$=1.1 nm and varying $w$ accordingly. In PWP simulation, we conducted three cases which are $d$:$w$=1:4, 1:3 and 1:2, and these area fraction are comparable with those in [S8]. We prescribe three different coherence lengths $C_l$=354, 601 and 1273 nm for each case to also study its dependence and $\alpha(\omega)$ results for LA phonons are shown in Fig. S3 (a). The resonant frequency of current configuration can be estimated based on Eq. (7) in [S8], which reads $\omega = v_{L,T} \sqrt{N_1^2 + N_2^2}/l$, where $v_{L,T}$ is the longitudinal or shear wave speed, $N_1$ and $N_2$ are the Miller indices for periodic square lattice at transverse direction, and $l$ is the length of square lattice. According to [S8], the resonance corresponding to the largest transmittance dip is interpreted as rigid-body resonance (rattling) of the spherical object, and the lattice resonance is excited as a result of shear waves propagating in the plane of the lattice. We estimated average shear wave speed of Si matrix from SW potential to be $v_T$=4534 m/s. The Miller indices at the largest transmittance dip are taken as {1 0} which includes four equivalent cases which are [1 0], [$\bar{1}$ 0], [0 1] and [0 $\bar{1}$] and $l$ equals to $w$. The resonant frequencies are extracted from Fig. S3(a) and plotted together with calculated frequencies by the analytical equation in Fig. S3(b). We also superimpose them with calculated eigenfrequencies from COMSOL Multiphysics® with additional case $d$:$w$=1:5, whose transmittance dip would become very shallow to discern in PWP simulation. A good mutual agreement of resonant frequencies among analytical equation, PWP simulation and COMSOL computation except for $d$:$w$=1:2 is observed, while the discrepancy between analytical equation and others with $d$:$w$=1:2 is possibly because, that close proximity of objects destroys the assumption of discrete lattice point implicitly made for the analytical equation.

Secondly, given the particle is not closely packed, i.e. its area faction is not too large, it is interesting to compare the analytical equation with PWP simulation with respect to the positions of transmittance dips, which corresponds to resonant scatterings originated from various Miller indices. The comparison is made with $d$:$w$=1:3 and the results are shown in Fig. S4. It shows general good agreement of resonant frequencies between PWP simulations and analytical solutions with several Miller indices, which are marked beside the corresponding transmittance dip.

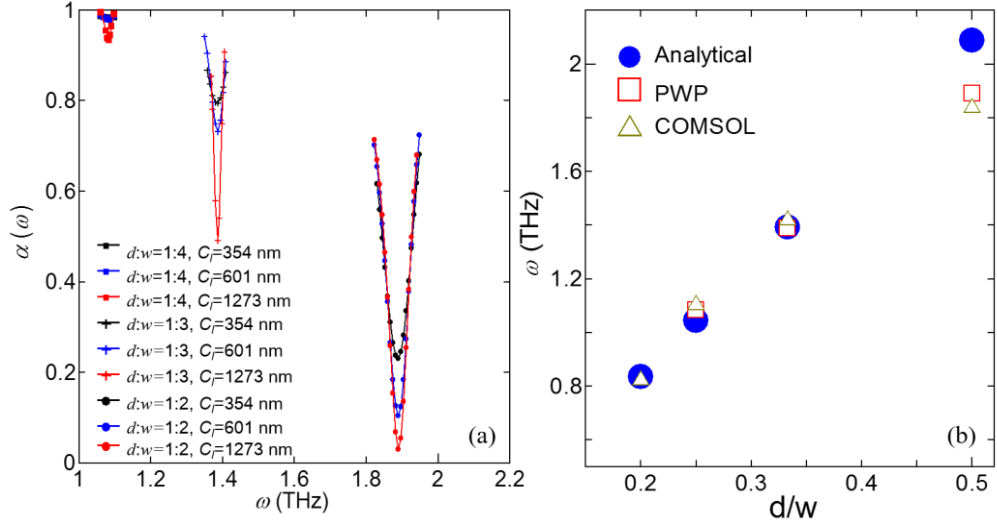

**Fig. S3** (a) $\alpha(\omega)$ in the vicinity of the largest transmittance dips calculated from LA PWP simulation for various ratios of $d$:$w$=1:4, 1:3 and 1:2 with different coherence lengths $C_l$=354, 601 and 1273 nm for each case, and $d$ is fixed to be 1.1 nm and $w$ is varied accordingly. (b) Dependence of resonant frequencies $\omega$ on $d$:$w$ ratios calculated from analytical equation, LA PWP simulation and COMSOL computation.

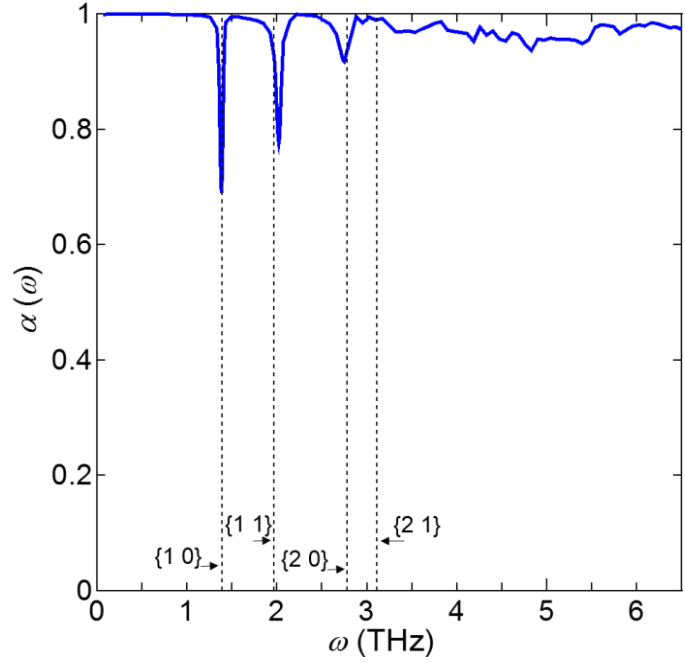

**Fig. S4** $\alpha(\omega)$ calculated from LA PWP simulation for $d:w$=1:3 with $d$=1.1 nm and $C_l$=601 nm. Four vertical dotted lines indicate resonant frequencies calculated from analytical equations with corresponding Miller indices marked in braces.

5) PWP coherence length dependence of the resonant transmittance dip width

We present PWP coherence length ($C_l$=85, 177, 354, 601 and 1273 nm) dependence of the largest resonant transmittance dip width by taking LA phonon with $d$=1.1 nm and $w$=2.2 nm as example in Fig. S5. The dip width is characterized by the full width at the half maximum (FWHM) of the transmittance dip and thus has the unit of THz. It is found that the FWHM generally decreases as $C_l$ increases and is thus inversely proportional to $C_l$ as demonstrated in [S9].

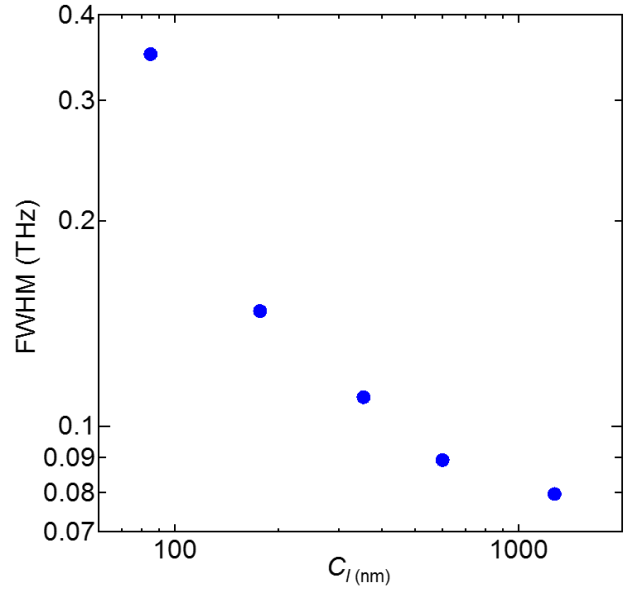

**Fig. S5** FWHM of the largest transmittance dip for LA PWP in the case of $d=1.1$ nm with $C_l=$85, 177, 354, 601 and 1273 nm.

6) Details of atomistic Green's function (AGF) method

The targeted system contains three coupled subsystems: two semi-infinite leads connected through the scattering region according to scattering theory as shown in Fig. S6.

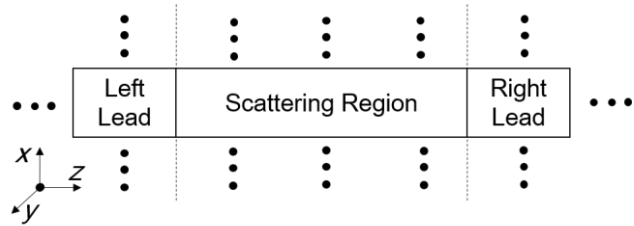

**Fig. S6** Schematics of system in AGF method.

The heat flux flowing in along the system axis is given by

$$J = \int_{BZ} \hbar\omega_\mathbf{k} v_{g,\mathbf{k}_z}(n_L - n_R) t_\mathbf{k} \frac{d^3\mathbf{k}}{(2\pi)^3}, \quad (S2)$$

where $\hbar\omega_\mathbf{k}$ is the energy of the phonon mode $\mathbf{k}$, $v_{g,\mathbf{k}_z}$ is the phonon group velocity of the phonon mode $\mathbf{k}_z$, $n_{L,R}$ is the phonon number on the left and right reservoir following the Bose-Einstein distribution $n = (\exp(\hbar\omega/k_B T)-1)^{-1}$, $t_\mathbf{k}$ is the transmission probability of the phonon mode $\mathbf{k}$. The integration goes through all the phonon modes in the full Brillouin Zone (BZ). In the linear regime, the phonon population undergoes small perturbations and thus the thermal conductance writes

$$G = J/\Delta T = \int_{BZ} \hbar\omega_\mathbf{k} v_{g,\mathbf{k}_z} \frac{\partial n}{\partial T} t_\mathbf{k} \frac{d^3\mathbf{k}}{(2\pi)^3}, \quad (S3)$$

We note that $d^3\mathbf{k} = dk_x dk_y dk_z$, and $v_{g,\mathbf{k}_z} dk_z = \partial\omega/\partial k_z dk_z = d\omega$. The above equation reduces to

$$G = J/\Delta T = \int_{BZ} \hbar\omega_\mathbf{k} \frac{\partial}{\partial T}(e^{\frac{\hbar w}{k_B T}} - 1)^{-1} [t_\omega dk_x dk_y] \frac{d\omega}{(2\pi)^3}. \quad (S4)$$

Hence we identify the spectral phonon transmission function $\Xi(\omega) = t_\omega g(\omega)$ where $g(\omega) = dk_x dk_y$ is the projected phonon density of states in the periodic directions of the system and $t_\omega$ is the transmission probability at frequency $\omega$.

We probe the spectral phonon transmission function $\Xi(\omega)$ by atomistic green's function (AGF) and the thermal conductance can be obtained by following the Landauer formula

[S10]:

$$G = \int_0^{\omega_{max}} \Xi(\omega) \frac{\partial}{\partial T}(e^{\frac{\hbar\omega}{k_B T}} - 1)^{-1} \hbar\omega \frac{d\omega}{2\pi}, \tag{S5}$$

where $\hbar\omega_k$ and $\omega_{max}$ are the energy and the maximum frequency in the system. $T$ refers to the mean temperature of the system, $k_B$ and $\hbar$ represent the Boltzmann and the reduced Planck constants, respectively. The transmission function $\Xi(\omega)$ is obtained from the atomistic Green's function approach as $Tr[\Gamma_L G_s \Gamma_R G_s^+]$. The advanced and retarded Green functions $G_s^+$ and $G_s$ can be deduced from:

$$G_s = [(\omega + i\Delta)^2 I - K_s - \Sigma_L - \Sigma_R]^{-1}, \tag{S6}$$

where $\Delta$ is an infinitesimal imaginary part that maintains the causality of the Green's function and $\Sigma_L = K_{ab} g_L K_{ab}^+$, $\Sigma_R = K_{ab} g_R K_{ab}^+$ are the self-energies of the left and right leads, the "+" exponent indicating the Hermitian conjugation. Finally, $g_L$ and $g_R$ refer to the surface Green's functions of the left and right leads, while $K_s$ and $K_{ab}$ are the force constant matrices derived from the potential, for the scattering region and between neighboring atoms in the leads, respectively. The expression of the transmission also includes broadening matrices, $\Gamma_L = i(\Sigma_L - \Sigma_L^+)$ and $\Gamma_R = i(\Sigma_R - \Sigma_R^+)$.

7) Consistency between atomistic Green's function (AGF) and phonon wave-packet (PWP) simulation on the resonant frequency and transmittance dip

We demonstrate the consistency of AGF with PWP regarding to the depiction of resonant transmittance dip for single GeNP with $d$=1.1 nm by setting transverse phonon wavevector $k_x=k_y=0$ (Γ-point), in accordance with that in PWP. As shown in the inset (i) of Fig. S7, there is simultaneous presence of both LA and TA phonon modes till about 2.2 THz in AGF calculation, while PWP deals with single mode within the same frequency range as shown in the inset (ii) and (iii). The agreement with respect to resonant frequencies for both LA and TA is thus as expected, but with different dip depths. The $α(ω)$ is expected to reach zero at resonant frequency with infinite coherence length assumed here in AGF calculation, but its magnitude for single mode should be weighted as a consequence of simultaneous presence of one LA mode and two TA modes below 2.2 THz. This leads to upper bound for $α_{max}(ω_{LA}) = 2/3$ and $α_{max}(ω_{TA}) = 1/3$ if only considering pure LA and TA resonance, respectively.

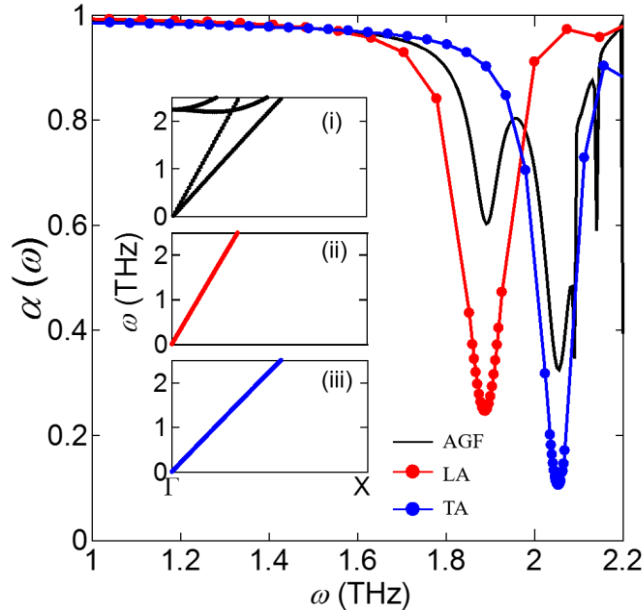

**Fig. S7** Comparison of $α(ω)$ calculated from AGF and PWP. Insets: phonon dispersion along Γ-X direction for that used in (i): AGF, (ii): LA PWP and (iii): TA PWP, respectively.

8) The collective resonance for TA modes

The TA mode generally shows very similar collective features as its LA counterpart as shown in Fig. S8 for four GeNPs array ($d$=1.1 nm) with $D=d$, $2d$, $4d$ and $5d$. When $D=2d$, because of the strong collective motion, the transmittance dip becomes the wideset among all cases and reach very close to zero. $D=4d$ shows similar but weaker broadening, a confirmation for the rule of integral multiples of $2d$. In the case of $D=5d$, the dip width is again narrower due to absence of collective resonance.

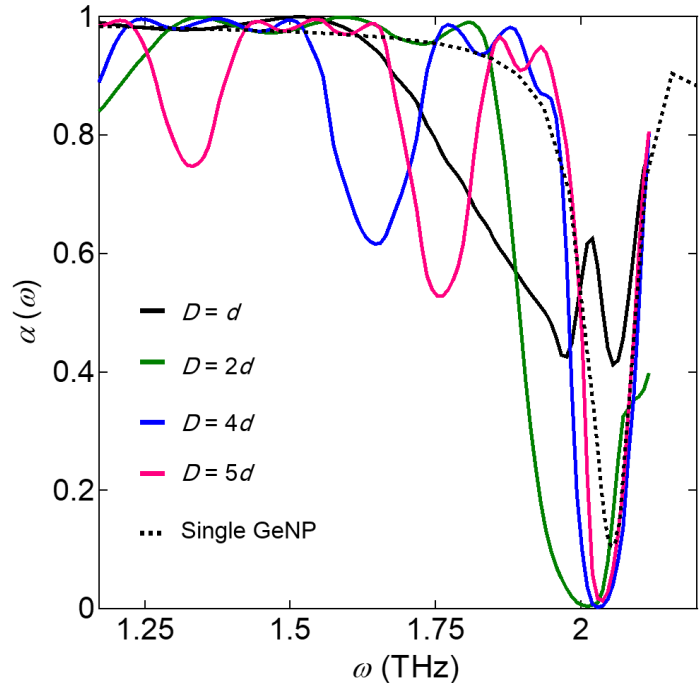

**Fig. S8** Transmittance $\alpha(\omega)$ calculated from TA PWP simulation with four GeNPs array ($d$=1.1 nm) with different equal inter-particle distances $D=d$, $2d$, $4d$ and $5d$.

9) Resonant squeeze mode

A large coherence length finds the presence of a secondary dip (Fig. 2 (e)), and we identify it is a resonant squeeze mode: We conducted the vibrational modes analyses to both the 1st (largest) and 2nd dip for the LA phonon at two different coherence lengths ($C_l$=177 nm and 1273 nm) as shown in Fig. S9. We found that on contrary to rattling mode in Fig. S9 (a) and (b) which shows a symmetric displacement (equal amplitudes at $z$+ and $z$- directions) of center of mass (COM) of GeNP along $z$-axis for 1st dip, it shows a nearly-zero motion for 2nd dip and an asymmetric feature in the displacement as shown in (e) and (f). The increasing $C_l$ increase the displacement for the 1st dip and enhance the asymmetric feature for the 2nd dip. We dig into more details by computing displacement of individual Ge atoms (total 34 atoms superimposed) comprising the GeNP along $x$-, $y$- and $z$-axes. As another confirmation for the rattling mode, it is found the Ge atoms vibrate along $z$-axis but nearly-zero motion along $x$- and $y$-axes [(c), (d)]. However, it shows displacements at all axes for the 2nd dip with amplitude along $z$-axis almost two times larger than those in $x$- and $y$-axes [(g), (h)]. Together with the vibrational behavior of GeNP in real space observed from PWP simulation, we conclude it is the alternative squeeze of GeNP along the $z$-axis and perpendicular to it for the 2nd dip.

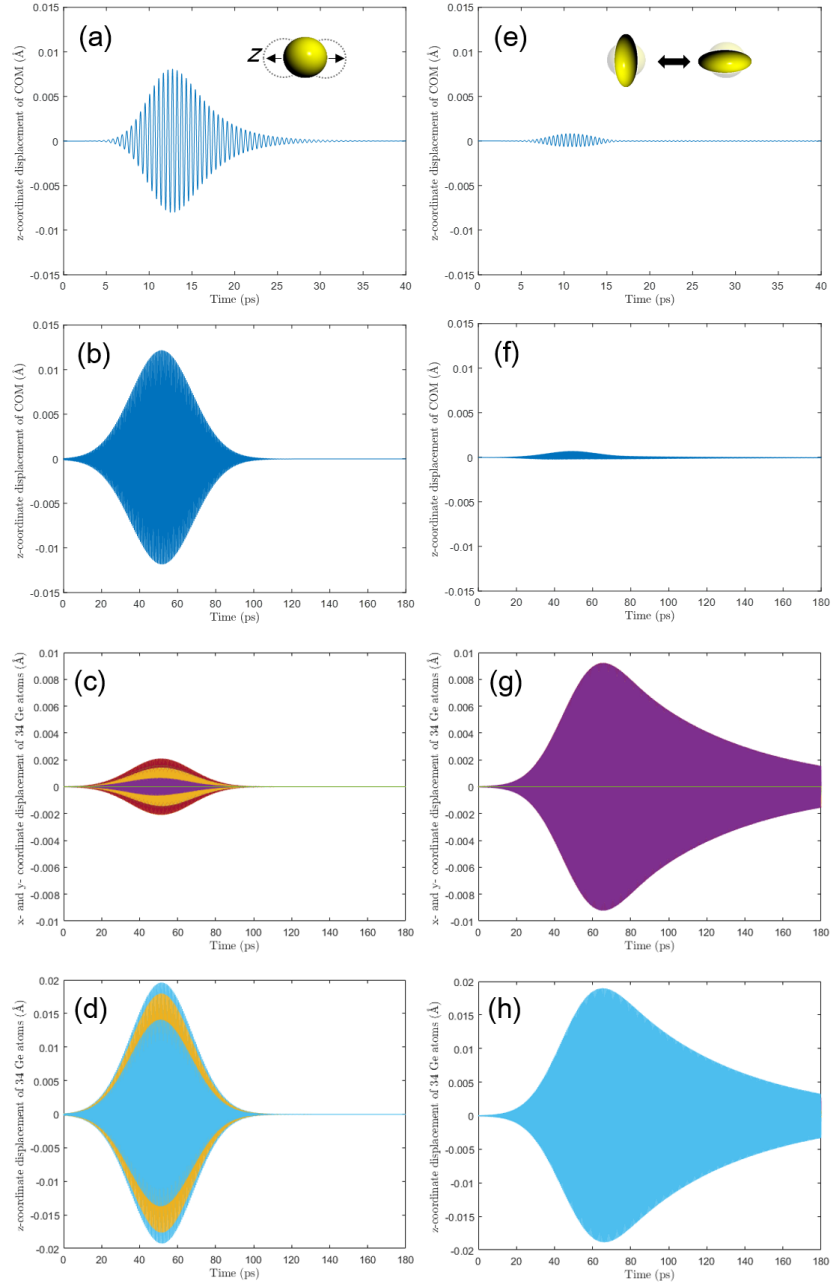

**Fig. S9** Time-evolution of $z$-coordinate displacements of center of mass (COM) of single GeNP ($d$=1.1 nm) corresponding to several frequencies of transmittance minima for LA phonons in Fig. 2(a). (a) The 1st (largest) dip at $C_l$=177 nm. Inset: schematic for the rattling mode. (b) The 1st dip at $C_l$=1273 nm. (e) The 2$^{nd}$ dip at $C_l$=177 nm. Inset: schematic for the squeeze mode. (f) The 2nd dip at $C_l$=1273 nm. Time-evolution of $x$-, $y$-

or *z*-coordinate displacements of individual Ge atoms (total 34 atoms) comprising the GeNP (*d*=1.1 nm) corresponding to several frequencies of transmittance minima for LA phonons in Fig. 2(a) with $C_l$=1273 nm. (c) *x*- and *y*-coordinate superimposed for 1st dip. (d) *z*-coordinate for 1st dip. (g) *x*- and *y*-coordinate superimposed for 2nd dip. (h) *z*-coordinate for 2nd dip.

## Reference

[S1] P. K. Schelling, S. R. Phillpot, and P. Keblinski, Phonon wave-packet dynamics at semiconductor interfaces by molecular-dynamics simulation, Applied Physics Letters **80**, 2484 (2002).

[S2] K. Naoaki, Y. Takahiro, and W. Kazuyuki, Phonon wavepacket scattering dynamics in defective carbon nanotubes, Japanese Journal of Applied Physics **45**, L963 (2006).

[S3] S. Ju and X. Liang, Detecting the phonon interference effect in Si/Ge nanocomposite by wave packets, Applied Physics Letters **106**, 203107 (2015).

[S4] W. Zhang, T. S. Fisher, and N. Mingo, Simulation of Interfacial Phonon Transport in Si–Ge Heterostructures Using an Atomistic Green's Function Method, Journal of Heat Transfer **129**, 483 (2007).

[S5] F. H. Stillinger and T. A. Weber, Computer simulation of local order in condensed phases of silicon, Physical Review B **31**, 5262 (1985).

[S6] T. Murakami, T. Hori, T. Shiga, and J. Shiomi, Probing and tuning inelastic phonon conductance across finite-thickness interface, Applied Physics Express **7**, 121801 (2014).

[S7] S. Plimpton, Fast Parallel Algorithms for Short-Range Molecular Dynamics, Journal of Computational Physics **117**, 1 (1995).

[S8] V. K. Kinra, K. Maslov, B. K. Henderson, N. Day, and G. Diderich, *Review of Progress in Quantitative Nondestructive Evaluation: Volume 16A*, pp. 51 (Springer US, Boston, MA, 1997).

[S9] Yu. A. Kosevich, H. Han, L. G. Potyomina, A. N. Darinskii, and S. Volz, *Quodons in Mica*, pp. 247 (Springer International Publishing Switzerland, Switzerland, 2015).

[S10] I. M. Khalatnikov, *An Introduction to the Theory of Superfludity* (Addisson-Wesley, 1989).